\documentclass[10pt,twocolumn,letterpaper]{article}

\usepackage{caption}
\usepackage{subcaption}
\usepackage{wacv}
\usepackage{times}
\usepackage{epsfig}
\usepackage{graphicx}
\usepackage{amsmath}
\usepackage{amssymb}
\usepackage{booktabs}
\usepackage{multirow} 
\usepackage{pgfplots}
\pgfplotsset{compat=1.12}
\pgfplotsset{ignore zero/.style={%
  #1ticklabel={\ifdim\tick pt=0pt \else\pgfmathprintnumber{\tick}\fi}
}}

\def\wacvPaperID{68} 

\wacvapplicationstrack 

\wacvfinalcopy 

\ifwacvfinal
\usepackage[breaklinks=true,bookmarks=false]{hyperref}
\else
\usepackage[pagebackref=true,breaklinks=true,colorlinks,bookmarks=false]{hyperref}
\fi

\renewcommand\sectionautorefname{Sec.}
\renewcommand\equationautorefname{Eq.}
\renewcommand\tableautorefname{Tab.}
\renewcommand\figureautorefname{Fig.}
\renewcommand\appendixautorefname{Appx.}

\begin{document}

\title{Brainomaly: Unsupervised Neurologic Disease Detection\\Utilizing Unannotated T1-weighted Brain MR Images
\vspace{-15pt}
}

\author{Md Mahfuzur Rahman Siddiquee\textsuperscript{1,2}, Jay Shah\textsuperscript{1,2}, Teresa Wu\textsuperscript{1,2}, Catherine Chong\textsuperscript{2,3},\\
Todd J. Schwedt\textsuperscript{2,3}, Gina Dumkrieger\textsuperscript{3}, Simona Nikolova\textsuperscript{3}, and Baoxin Li\textsuperscript{1,2}\\
\textsuperscript{1}Arizona State University; \textsuperscript{2}ASU-Mayo Center for Innovative Imaging; \textsuperscript{3}Mayo Clinic
}


\twocolumn[{%
\renewcommand\twocolumn[1][]{#1}%
\maketitle
\vspace{-170pt}
\begin{center}
    \centering
    \textcolor{blue}{Please cite the paper as M.M. Rahman Siddiquee, J. Shah, T. Wu, C. Chong, T. J. Schwedt, G. Dumkrieger, S. Nikolova, and B. Li. Brainomaly: Unsupervised Neurologic Disease Detection Utilizing Unannotated T1-weighted Brain MR Images. Proceedings of the IEEE/CVF Winter Conference on Applications of Computer Vision (WACV), 2024.}
\end{center}%
\vspace{106pt}
}]

\begin{abstract}
Harnessing the power of deep neural networks in the medical imaging domain is challenging due to the difficulties in acquiring large annotated datasets, especially for rare diseases, which involve high costs, time, and effort for annotation. Unsupervised disease detection methods, such as anomaly detection, can significantly reduce human effort in these scenarios. While anomaly detection typically focuses on learning from images of healthy subjects only, real-world situations often present unannotated datasets with a mixture of healthy and diseased subjects. Recent studies have demonstrated that utilizing such unannotated images can improve unsupervised disease and anomaly detection. However, these methods do not utilize knowledge specific to registered neuroimages, resulting in a subpar performance in neurologic disease detection. To address this limitation, we propose Brainomaly, a GAN-based image-to-image translation method specifically designed for neurologic disease detection. Brainomaly not only offers tailored image-to-image translation suitable for neuroimages but also leverages unannotated mixed images to achieve superior neurologic disease detection. Additionally, we address the issue of model selection for inference without annotated samples by proposing a pseudo-AUC metric, further enhancing Brainomaly's detection performance. Extensive experiments and ablation studies demonstrate that Brainomaly outperforms existing state-of-the-art unsupervised disease and anomaly detection methods by significant margins in Alzheimer's disease detection using a publicly available dataset and headache detection using an institutional dataset. The code is available from \href{https://github.com/mahfuzmohammad/Brainomaly}{https://github.com/mahfuzmohammad/Brainomaly}.
\end{abstract}

\begin{figure}[!htp]
    \centering
    \includegraphics[trim={0.3cm 0 0.3cm 0cm},clip,width=\linewidth]{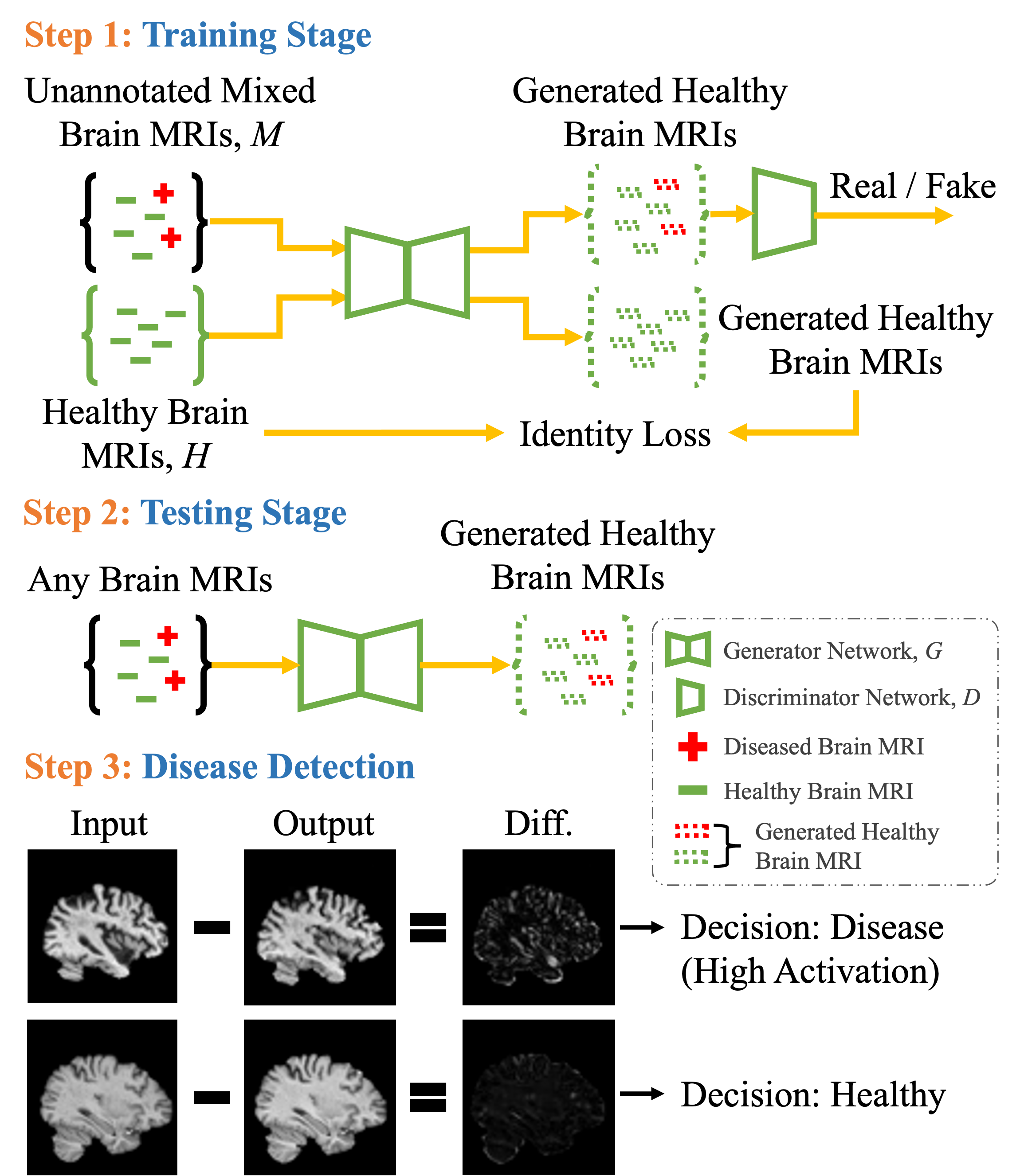}
    \caption{Overview of the proposed method, Brainomaly, for unsupervised neurologic disease detection using unannotated mixed T1-weighted brain MRIs. Brainomaly is a GAN-based image-to-image translation method that is trained (Step~1~in the figure) to remove the diseased regions from any input brain MRI and generate MRI of the corresponding healthy brain using (1) a set of ``unannotated mixed brain MRIs'' containing T1-weighted brain MRIs from individuals with neurologic disease as well as healthy subjects and (2) another set containing T1-weighted brain MRIs only from healthy subjects. Once trained, the generator turns any brain MRI into the MRI of the corresponding healthy brain (Step~2~in the figure). Hence, subtracting (Step~3~in the figure) the generated MRI of the healthy brain from its input would reveal structural changes if the input MRI is of an abnormal brain. We use the average value of the resultant difference map as the disease detection score, where higher values indicate a higher likelihood that the brain MRI is from someone with a neurologic disease.}
    \label{fig:anomaly_detection_overview}
    \vspace{-10pt}
\end{figure}

\vspace{-10pt}

\section{Introduction}
\label{sec:introduction}

Deep neural networks have facilitated supervised learning from annotated datasets~\cite{he2015delving,esteva2017dermatologist}, but acquiring large annotated medical imaging datasets, particularly for rare diseases, is challenging. Even when enough imaging data are available, manual annotation of such datasets is expensive, laborious, and time-consuming as it requires domain expert knowledge. In such scenarios, unsupervised disease detection methods like anomaly detection can help reduce the annotation burden and save significant human effort. Many prior works in this area have focused on developing diagnostic models that learn to reconstruct images from healthy subjects~\cite{chen2018unsuperviseddetection,schlegl2017unsupervised,schlegl2019fdetection,alex2017generative,zenati2018efficientdetection,zenati2018adversarially,akcay2019ganomaly,gherbi2019encodingdetection,sabokrou2018adversariallydetection}. These methods rely on poor reconstructions of images from individuals with diseases during inference for detection. However, in practice, {\em unannotated} images are often available from individuals with diseases (mixed with healthy subject images) in clinical databases or even as test sets where the trained model will be applied, and leveraging the additional information contained in these {\em unannotated mixed} images could enhance disease detection.

With the same inspiration,~\cite{rahman2022healthygan} and~\cite{cai2022dual} have recently proposed methods that utilize \textit{such} unannotated datasets of mixed images during training for improved unsupervised patient-level disease and anomaly detection.~\cite{cai2022dual} trained a set of autoencoders to reconstruct chest X-ray images only from healthy subjects and another set of autoencoders to reconstruct unannotated mixed X-rays from individuals with thoracic diseases and healthy subjects. Anomaly detection scores were obtained by comparing the discrepancies between these two sets of autoencoders. Conversely,~\cite{rahman2022healthygan} employed a GAN-based image-to-image translation approach~\cite{goodfellow2014generative,goodfellow2020generative,isola2017image,choi2018stargan,choi2020stargan,siddiquee2019learning} to remove diseased areas from input images and generate corresponding healthy-looking images. The disease detection scores for each subject were calculated by subtracting the generated healthy-looking images from their corresponding input images. These methods, however, performed suboptimally for neurologic disease detection (\tableautorefname~\ref{tab:detection_results}) and lacked a reliable inference model selection criterion.~\cite{cai2022dual} used the model from the last training iteration for inference while~\cite{rahman2022healthygan} selected model based on the realism of the generated images using Fréchet inception distance (FID)~\cite{heusel2017gans}. We found that the FID metric has a weak correlation with the underlying classification performance of the model (see~\sectionautorefname~\ref{subsec:results_abl}). To address these issues, we propose {\em Brainomaly}, a GAN-based image-to-image translation method specifically designed for neurologic disease detection. Brainomaly learns to remove neurologic disease from T1-weighted brain MRIs and generates corresponding healthy MRIs. During training, it utilizes an unannotated set of mixed MRIs from diseased and healthy individuals where traditional cycle-consistency-based image translation is not applicable~\cite{rahman2022healthygan}. Since neuroimages are usually registered, we design Brainomaly to predict an {\em additive map} to transform input images into a healthy appearance instead of directly generating healthy images. The additive map contains voxel-wise values representing the estimated changes required to transform the input MRI into a healthy brain. We hypothesize that this additive map-based translation, combined with {\em identity loss} (\equationautorefname~\ref{eq:id_loss}) regularization, relaxes the need for cycle-consistency-based image translation. For inference model selection, we introduce a {\em pseudo-AUC} (AUCp) metric that further boosts the detection performance.~\figureautorefname~\ref{fig:anomaly_detection_overview} depicts an overview of the Brainomaly framework.

Through extensive experiments and ablation studies, we demonstrate that Brainomaly outperforms existing state-of-the-art unsupervised disease and anomaly detection methods by significant margins on one public dataset for Alzheimer's disease detection and one institutional dataset for headache detection. Its superior performance is due to the additive map-based image translation technique, leveraging unannotated images during training and employing improved inference model selection using AUCp. In summary, we make the following contributions:
\begin{itemize}
    \item We introduce a novel neurologic disease detection method that utilizes unannotated T1-weighted brain MRIs from individuals with neurologic disease and healthy subjects.
    \item We propose a new metric, AUCp, for selecting a suitable model for inference when an annotated validation dataset is unavailable.
    \item With two neuroimaging datasets, we perform extensive experiments comparing the proposed method, Brainomaly, against the conventional state-of-the-art unsupervised patient-level disease and anomaly detection methods. Our detailed analysis proves the superiority of the proposed method over existing methods.
    \item We evaluate the proposed method in both transductive and inductive settings to match real-world scenarios.
    \item We empirically show that our proposed metric has a higher correlation with the models' underlying disease detection performances and selects a higher-performing model than a model selected by FID, which is commonly used in GAN model development.
\end{itemize}

\section{Related Work}
\label{sec:related_works}

Our work is closely related to image-to-image translation, GAN-based anomaly detection, and neurologic disease detection. Hence, we review relevant existing efforts on these tasks and contrast them with our proposed neurologic disease detection method, Brainomaly.

\subsection{Image-to-Image Translation}


Plenty of work has been done on GAN-based image-to-image translation~\cite{isola2017image,kim2017learning,ledig2017photo,liu2017unsupervised,shen2017learning,yi2017dualgan,zhu2017unpaired,zhu2017toward,choi2018stargan,alami2018unsupervised,zhang2018generativetrans,he2019attgantrans,liu2019stgan,nizan2020breaking}. Pix2Pix~\cite{isola2017image} and CycleGAN~\cite{zhu2017unpaired} are pioneer methods in this area. While Pix2Pix requires paired input-output images for training, CycleGAN introduces the concept of cycle consistency for unpaired image-to-image translation. However, we cannot directly use CycleGAN in our work on unsupervised disease and anomaly detection due to the lack of annotated diseased images.


Recent unpaired image-to-image translation methods~\cite{liu2017unsupervised,shen2017learning,yi2017dualgan,zhu2017unpaired,zhu2017toward,choi2018stargan,alami2018unsupervised,zhang2018generativetrans,he2019attgantrans,liu2019stgan,zhao2020unpaired}, regardless of cycle-consistency, also rely on image annotations. For example,~\cite{shen2017learning} utilizes two generators for translating images of human faces between a pair of facial attributes.~\cite{alami2018unsupervised} proposes an attention-based approach that performs image-to-image translation like CycleGAN with two additional networks for generating attention maps. Alternatively, StarGAN~\cite{choi2018stargan}, AttGAN~\cite{he2019attgantrans}, STGAN~\cite{liu2019stgan}, and Fixed-Point GAN~\cite{siddiquee2019learning} utilize one generator network that requires target image annotations. A recent ensemble-based method~\cite{nizan2020breaking} offers an alternative to cycle consistency but demands multiple generator and discriminator networks, making it computationally expensive and difficult to train.


In contrast, our Brainomaly method, which employs only one generator and one discriminator network, overcomes the need for cycle consistency by generating additive maps instead of images. Furthermore, our Brainomaly approach performs better than existing anomaly detection methods, as shown in~\sectionautorefname~\ref{sec:results}.

\subsection{Anomaly Detection}
\label{subsec:related_anomaly}


In general, GAN-based anomaly detection methods~\cite{chen2018unsuperviseddetection,schlegl2017unsupervised,schlegl2019fdetection,alex2017generative,zenati2018efficientdetection,zenati2018adversarially,akcay2019ganomaly,gherbi2019encodingdetection, cai2022dual} in the existing literature primarily focus on learning from {\em healthy} images. These methods aim to capture the underlying manifold of healthy images, enabling their decoders to reconstruct only healthy images during testing. Consequently, when {\em diseased} images are reconstructed as healthy, the disparity between the input and output images indicates the presence of anomalies. We elaborate on a few examples below.



Chen \etal~\cite{chen2018unsuperviseddetection} employ an adversarial autoencoder to learn the distribution of healthy data. They identify anomalies by subtracting the reconstructed diseased image from the input image. In a similar vein,~\cite{schlegl2017unsupervised} proposes a method that adversarially learns a decoder model to generate healthy images from random noise vectors in the latent space. During testing, this approach maps new images to the latent space through iterative updates of the latent vector. If a new image is healthy, the method is expected to find the actual latent vector that reconstructs the input image, resulting in a negligible difference between the input and reconstructed images. Conversely, for diseased images, the method should find a latent vector that produces the closest healthy image, leading to a higher difference between the input and reconstructed images. The authors propose an anomaly score that combines the reconstruction error and the discrimination score from the discriminator network.


Schlegl \etal~\cite{schlegl2019fdetection} enhance the speed of~\cite{schlegl2017unsupervised} by introducing an encoder network capable of mapping input images to the latent space in a single pass. Likewise,~\cite{alex2017generative} employs a GAN to learn a generative model of healthy data. It involves scanning images pixel-by-pixel and feeding the cropped regions to a trained GAN discriminator. An anomaly map is then constructed by combining the anomaly scores provided by the discriminator. Zenati \etal~\cite{zenati2018efficientdetection,zenati2018adversarially} utilize BiGAN~\cite{donahue2016adversarialdetection} to jointly train an encoder and a decoder network to learn the mapping of normal images. Like most methods, they use the reconstruction error as the anomaly score.

Akcay \etal~\cite{akcay2019ganomaly} train an autoencoder supervised with both image-level $L_1$ distance and adversarial loss using only normal images. Additionally, they train an extra encoder to map the images reconstructed by the autoencoder back to their latent space. In a different approach,~\cite{gherbi2019encodingdetection} trains an encoder network to map normal images to a Gaussian distribution and abnormal images to an out-of-distribution region using adversarial learning. Anomalies are then detected using the Mahalanobis distance in the latent space. It is important to note that this method requires annotated anomalous images during training.


In contrast,~\cite{rahman2022healthygan} and~\cite{cai2022dual} learn from unannotated images of both diseased and healthy individuals, similar to our proposed method as discussed in~\sectionautorefname~\ref{sec:introduction}. However, in~\sectionautorefname~\ref{sec:results}, we demonstrate that our method significantly outperforms these existing methods by a large margin.

\subsection{Neurologic Disease Detection}

Numerous studies have explored deep learning techniques for automated Alzheimer's disease diagnosis using raw imaging data. Most of these studies focus on supervised classification tasks, while a few employ unsupervised anomaly detection methods~\cite{cabreza2022anomaly, han2021madgan, baydargil2021anomaly, jin2021unsupervised, bai2022novel, choi2019deep}. For instance, Cabreza \etal~\cite{cabreza2022anomaly} train a GAN on healthy images, followed by an encoder that returns a vector for input images like~\cite{schlegl2019fdetection}. MADGAN~\cite{han2021madgan} leverages MRI slice continuity in reconstruction and uses high reconstruction loss for anomalous image classification. Baydargil \etal~\cite{baydargil2021anomaly} incorporate a parallel feature extractor within a GAN using PET images, while Choi \etal~\cite{choi2019deep} employ a variational autoencoder on PET images for abnormality scoring based on reconstruction error. Jin \etal~\cite{jin2021unsupervised} use an adversarial autoencoder for unsupervised data characterization of healthy controls and subsequent Alzheimer's disease vs. healthy control classification. Bai \etal~\cite{bai2022novel} combine a classifier with GAN training, incorporating high-level feature extraction and posterior class probabilities.


Except for~\cite{bai2022novel}, the aforementioned approaches rely solely on healthy images for GAN training and utilize high reconstruction loss to detect anomalies. In contrast, our method leverages both unannotated mixed images and healthy images during training, leading to superior Alzheimer's disease detection than state-of-the-art methods.


Regarding headache detection from structural MRI scans, we found no unsupervised approaches in the literature. Only a few studies employ deep learning techniques for this task. Rahman Siddiquee \etal~\cite{rahman2023headache} develop a ResNet-based binary classification model for automated biomarker extraction of headache sub-types. Yang \etal~\cite{yang2018multimodal} proposes a deep convolutional neural network using pre-processed resting-state fMRI data to distinguish between migraine and healthy controls. However, both studies are supervised classification tasks and suffer from limited datasets, which is common in headache classification using deep learning. Thus, our method enhances unsupervised headache detection by utilizing unannotated MRIs from both headache and healthy subjects.

\section{The Proposed Method: Brainomaly}
\label{sec:method}

\figureautorefname~\ref{fig:anomaly_detection_overview} depicts the overview of the proposed method. This section presents details about the neural network models, their training process, selecting the inference model with the proposed AUCp metric, and neurologic disease detection using the outputs from these models.

\subsection{The Networks}

Brainomaly consists of a discriminator network and a generator network. The discriminator network follows PatchGAN~\cite{isola2017image,li2016precomputed,zhu2017unpaired} architecture and is similar to the ones used in~\cite{choi2018stargan,siddiquee2019learning,rahman2022healthygan}. Our discriminator distinguishes whether a T1-weighted brain MRI is real or generated.

The generator network is an encoder-decoder type architecture similar to the generator networks used in~\cite{choi2018stargan,siddiquee2019learning,rahman2022healthygan}. As input, the network takes a T1-weighted brain MRI of any subject without knowing whether the subject is healthy or has a neurologic disease. As output, it generates an {\em additive map} where each voxel contains the value of an estimated required change to turn the brain in the input MRI into a healthy brain. The final healthy-brain MRI is generated by first summing the input MRI and the additive map voxel-wise, then applying $tanh$ activation on the resultant.

Both our generator and discriminator networks operate on 2D MRI slices. The generated healthy-brain MRI is constructed by stacking the generated MRI slices as they appeared in the input MRI. The architecture details for both these networks are provided in the~\appendixautorefname~\ref{appx:arch}.

\begin{figure*}[t!]
    \centering
    \includegraphics[trim={0.4cm 0 0cm 0.4cm},clip,width=\linewidth]{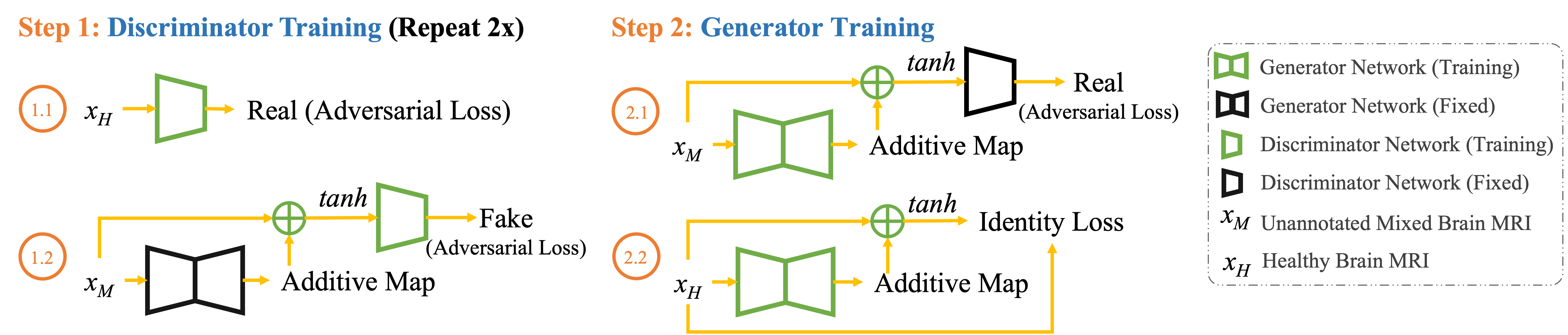}
    \caption{A training iteration in Brainomaly involves two steps: (1) the training of the discriminator and (2) the training of the generator. The discriminator is taught to identify the difference between real images and those generated by the generator. Meanwhile, the generator attempts to create indistinguishable images from real images to trick the discriminator. The iteration is repeated multiple times during the training phase to improve the realism of the generated images.}
    \label{fig:healthy_gan_training}
    \vspace{-10pt}
\end{figure*}

\subsection{Training the Networks}
\label{subsec:method_training}

\figureautorefname~\ref{fig:healthy_gan_training} provides a detailed schema for training the generator and discriminator networks of Brainomaly. We train the generator and the discriminator network alternately, like any GAN model. At each training step, we update the generator's weights once for every two weight updates of the discriminator network and repeat them until convergence.

The role of the discriminator network is to improve the generator network by providing iterative feedback during training. At each iteration (Step 1 in~\figureautorefname~\ref{fig:healthy_gan_training}), the discriminator network learns to distinguish the real MRIs of healthy brains and the generated MRIs from the previous iteration's generator. During the generator training, it provides feedback so that the generator can improve the quality of the generated MRIs. To be able to perform this role, the discriminator is trained on a set of T1-weighted healthy brain MRIs, $H$ (\figureautorefname~\ref{fig:anomaly_detection_overview}), to classify them as {\em real}, as well as on generated MRI slices to classify them as {\em fake}. During discriminator training, the generator solely uses MRIs from the unannotated mixed set $M$ (\figureautorefname~\ref{fig:anomaly_detection_overview}) to generate MRIs of healthy brains, excluding any MRIs from set $H$ as MRIs of healthy brains are already present in $M$. The training objective is achieved using an {\em adversarial loss} (\equationautorefname~\ref{eq:adv_d_loss_naive}).

\vspace{-0.3cm}
\begin{equation}
\label{eq:adv_d_loss_naive}
\begin{split}
\mathcal{L}_{adv}^{D} = \thinspace & ~\mathbb{E}_{x_M \in M}[\\
&log(1-D_{real/fake}(tanh(x_M+G(x_M))))]\\
&+ \mathbb{E}_{x_H \in H}[log(D_{real/fake}(x_H))]
\end{split}
\end{equation}

\noindent Here, $x_M$ and $x_H$ are MRI of random subjects from $M$ and $H$, respectively. The generator's output, which is the generated MR images of a healthy brain, is denoted as $G(.)$. The discriminator network's output is represented by $D_{real/fake}(.)$. We revised~\equationautorefname~\ref{eq:adv_d_loss_naive} based on the Wasserstein GAN~\cite{arjovsky2017wasserstein} and added a gradient penalty~\cite{gulrajani2017improved} with weight ${\lambda}_{gp}$ to enhance training stability. The revised training objective is shown in~\equationautorefname~\ref{eq:adv_d_loss} where $\hat{x}$ is a random weighted average of a batch of real and generated MRIs.

\vspace{-0.5cm}
\begin{equation}
\label{eq:adv_d_loss}
\begin{split}
\mathcal{L}_{adv}^{D} = \thinspace & ~\mathbb{E}_{x_M \in M}[D_{real/fake}(tanh(x_M + G(x_M)))]\\
&- \mathbb{E}_{x_H \in H}[D_{real/fake}(x_H)]\\
&+ {\lambda}_{gp} \thinspace {\mathbb{E}}_{\hat{x}}[{{(||{\triangledown}_{\hat{x}} {D}_{real/fake}(\hat{x})||}_{2} - 1)}^{2}]
\end{split}
\end{equation}

On the other hand, the generator aims to generate realistic MRIs of healthy brains by utilizing the discriminator's iterative feedback (Step 2 in~\figureautorefname~\ref{fig:healthy_gan_training}). For training, it translates MRI from both $M$ and $H$ sets and updates the generated MRIs of healthy brains iteratively and gradually so that the discriminator fails to distinguish the generated MRIs from the real ones. For the set of unannotated mixed MRIs ($M$), if the input is an MRI containing a neurologic disease, then the generator is expected to remove the diseased regions and generate the MRI of a corresponding healthy brain. If the input is already an MRI of a healthy brain, the generator is expected to behave like an autoencoder; that is, it is expected to generate exactly the same input MRI in output. As MRIs in set $M$ are unannotated, we used the {\em adversarial loss} for generating the corresponding MRIs of healthy brains. The loss is defined as in~\equationautorefname~\ref{eq:adv_g_loss}.

\vspace{-0.5cm}
\begin{equation}
\label{eq:adv_g_loss}
    \mathcal{L}_{adv}^{G} = - \mathbb{E}_{x_M \in M}[D_{real/fake}(tanh(x_M + G(x_M)))]
\end{equation}

\noindent For the set of healthy-brain MRIs ($H$), we explicitly train the generator to be an autoencoder using the {\em identity loss} (defined in~\equationautorefname~\ref{eq:id_loss}) for these MRI slices.

\vspace{-0.5cm}
\begin{equation}
\label{eq:id_loss}
    \mathcal{L}_{id} = \mathbb{E}_{x_H \in H}[||tanh(x_H + G(x_H)) - x_H||_1]
\end{equation}

Combining all these losses, the final full objective function for the discriminator and generator can be described by~\equationautorefname~\ref{eq:d_full_loss} and~\equationautorefname~\ref{eq:g_full_loss}, respectively.

\vspace{-0.2cm}
\begin{equation}
\label{eq:d_full_loss}
    \mathcal{L}_D = \mathcal{L}_{adv}^{D}
\end{equation}
\begin{equation}
\label{eq:g_full_loss}
    \mathcal{L}_G = \mathcal{L}_{adv}^{G} + \lambda_{id}\mathcal{L}_{id}
\end{equation}
where $\lambda_{id}$ is the relative importance of the \textit{identity loss}.

\subsection{Detecting the Diseases}
\label{subsec:method_disease_detection}

We detect diseases from the MRIs using {\em disease detection scores}. To get these scores, we translate all the given brain MRIs to MRIs of healthy brains using a trained generator model of Brainomaly (Step 2 in~\figureautorefname~\ref{fig:anomaly_detection_overview}). Then, we subtract the generated MRIs of healthy brains from their corresponding input MRIs (Step 3 in~\figureautorefname~\ref{fig:anomaly_detection_overview}). If the brain in the input MRI is diseased ({\em i.e.} abnormal), the resultant difference map would reveal structural changes. The difference map should reveal less or no structural changes for an input MRI of a healthy brain. We call the voxels showing the structural changes as {\em activations}. The average of all the activations in the difference map of an input MRI is its {\em disease detection score}, where a higher score indicates a higher likelihood of the input brain being diseased.

\subsection{Inference Model Selection using AUCp}
\label{subsec:method_validation}

As discussed in~\sectionautorefname~\ref{subsec:method_training} and~\figureautorefname~\ref{fig:healthy_gan_training}, Brainomaly learns iteratively from the given T1-weighted brain MRIs, generating multiple model checkpoints each after a fixed number of iterations. In a supervised learning setting, these models would be evaluated on a small validation dataset, and the best-performing model would be selected for inference and disease detection. However, annotated validation dataset is unavailable in our problem setting. Therefore, we use our proposed AUCp metric for model selection. To calculate AUCp, we first generate the {\em disease detection scores} for each model, as discussed in~\sectionautorefname~\ref{subsec:method_disease_detection}. As we already know that the set $H$ contains only healthy-brain MRIs, we assume the labels for MRIs in the unannotated mixed brain MRI set, $M$, to be diseased brains. Then, we use these {\em imperfect} annotations as ground truths along with the {\em disease detection scores} in the traditional AUC calculation resulting in AUCp scores. Once the AUCp scores are available for all the models, we select a model with the highest AUCp score for inference. In~\sectionautorefname~\ref{subsec:results_abl}, we have also shown that AUCp sets a better-performing model for inference compared to FID, commonly used in existing works~\cite{rahman2022healthygan}.~\appendixautorefname~\ref{appx:aucp} shows a schematic diagram for the AUCp calculation.

\section{Datasets}

\subsection{Alzheimer's Disease Dataset}

The Alzheimer's disease dataset was obtained from the ADNI database \href{http://adni.loni.usc.edu/}{(adni.loni.usc.edu)}, which is a large-scale public repository of clinical, neuropsychological, behavioral, genetic, and neuroimaging data to track the progression of Alzheimer's disease dementia. Using data from 3 studies that ADNI offers, ADNI-1, ADNI-2, and ADNI-GO, we collected and processed T1 MRI scans of 536 Alzheimer's disease  patients{\iffalse, 2606 mild cognitive impairment (MCI) patients,\fi} and 1271 healthy controls.

We randomly selected 501 MRIs from healthy controls for our experiments for the healthy brain MRI set ($H$). We created two unannotated mixed brain MRI sets ($M$): {\em AD DS1} and {\em AD DS2}. Each contains 268 MRIs from patients with Alzheimer's disease and 385 MRIs from healthy controls. Splitting the dataset helps us evaluate the proposed Brainomaly for Alzheimer's disease detection in both transductive and inductive settings in~\sectionautorefname~\ref{subsec:results_abl}.

All 3D MRIs in this dataset were registered to the {\em MNI152 1mm} template and skull stripped. We converted the 3D MRIs and saved them as 2D sagittal slices. The proposed Brainomaly method performs a prediction for each 2D slice. We aggregated the slice-level predictions by averaging them for patient-level predictions during evaluation.

\subsection{Headache Dataset}

We collected MRIs of 96 individuals with migraine, 48 with acute post-traumatic headache (APTH), 49 with persistent post-traumatic headache (PPTH), diagnosed according to the International Classification of Headache Disorders (ICHD) diagnostic criteria, and 104 healthy controls from Mayo Clinic. We extended our dataset by including MRIs of 428 healthy controls from the publicly available IXI dataset~\cite{ixi}. 

For our experiments, we trained our model by combining all headache types into one group first and then investigated each subgroup's performance separately in the post-analysis. We randomly selected 232 MRIs of healthy controls for the healthy brain MRI set ($H$). We created two unannotated mixed brain MRI sets ($M$): {\em HEAD DS1} and {\em HEAD DS2}. Each contains an equal number of MRIs for migraine (n = 48), APTH (n = 24), and healthy controls (n = 150). 24 out of 49 MRIs of those with PPTH were included in {\em HEAD DS1} and the rest in {\em HEAD DS2}. Similar to the experiment on Alzheimer's disease's dataset, such splitting helps evaluate the proposed Brainomaly for headache detection in both transductive and inductive settings in~\sectionautorefname~\ref{subsec:results_abl}.

All 3D MRIs in this dataset were registered to the {\em MNI152 1mm} template and skull stripped. We converted the 3D MRIs and saved them as 2D sagittal slices. We aggregated the slice-level predictions by averaging them for patient-level predictions during evaluation.

\begin{table*}[!htp]
    \centering
    \begin{tabular}{l | l | c | c | c | c | c | c}
        \hline
        \multirow{2}{*}{\textbf{Training Data}} & \multirow{2}{*}{\textbf{Methods}} & \multicolumn{3}{c |}{\textbf{Alzheimer's Disease Dataset}} & \multicolumn{3}{c}{\textbf{Headache Dataset}} \\
        \cline{3-8}
         & & {\em AD DS1} & {\em AD DS2} & \textbf{{\em Average}} & {\em HEAD DS1} & {\em HEAD DS2} & \textbf{{\em Average}}\\
        \hline
        \multirow{4}{*}{Healthy Only} & ALAD~\cite{zenati2018adversarially} & 0.5329 & 0.5239 & 0.5284 & \underline{0.7819} & 0.7486 & 0.7653 \\
        & ALOOC~\cite{sabokrou2018adversariallydetection} & 0.4670 & 0.4746 & 0.4708 & 0.3044 & 0.6566 & 0.4805 \\
        & f-AnoGAN~\cite{schlegl2019fdetection} & \underline{0.5946} & \underline{0.6093} & \underline{0.6020} & 0.4354 & 0.3925 & 0.4071 \\
        & Ganomaly~\cite{akcay2019ganomaly} & 0.5864 & 0.6048 & 0.5956 & 0.7313 & 0.6514 & 0.6913 \\
        \hline
        \multirow{5}{*}{Unannotated Mixed} & DDAD~\cite{cai2022dual} & 0.5897 & 0.5955 & 0.5926 & 0.6128 & 0.6431 & 0.6280 \\
        & HealthyGAN~\cite{rahman2022healthygan} & 0.4598 & 0.5468 & 0.5033 & 0.7107 & \underline{0.8333} & \underline{0.7720} \\
        & HealthyGAN (AUCp) & 0.5905 & 0.6034 & 0.5970 & 0.8276 & 0.7899 & 0.8088 \\
        & Brainomaly (FID) & 0.6389 & 0.6453 & 0.6421 & 0.9002 & 0.8589 & 0.8796 \\
        & \textbf{Brainomaly (AUCp)} & \textbf{0.6452} & \textbf{0.6648} & \textbf{0.6550} & \textbf{0.9041} & \textbf{0.8878} & \textbf{0.8960} \\
        \hline
    \end{tabular}
    \caption{Comparison of Brainomaly's performance with state-of-the-art anomaly detection methods on Alzheimer's disease and headache detection on unseen test sets using AUC metric. Numbers in \textbf{boldface} indicate the best results, and \underline{underlined} numbers indicate the second-best results. As seen, Brainomaly outperforms all the existing state-of-the-art methods for neurologic disease detection. The rows ``HealthyGAN (AUCp)'' and ``Brainomaly (FID)'' are for ablation study purpose only (see~\sectionautorefname~\ref{subsec:results_abl}).}
    \label{tab:detection_results}
\end{table*}

\section{Experiments}
\label{sec:experiments}

We evaluated our proposed Brainomaly on Alzheimer's disease (\sectionautorefname~\ref{subsec:ad_results}) and headache (\sectionautorefname~\ref{subsec:headache_results}) detection comparing with six state-of-the-art unsupervised disease/anomaly detection methods. Among these, DDAD~\cite{cai2022dual} and HealthyGAN~\cite{rahman2022healthygan} also utilize unannotated mixed images like Brainomaly. On the other hand, ALAD~\cite{zenati2018adversarially}, ALOOC~\cite{sabokrou2018adversariallydetection}, f-AnoGAN~\cite{schlegl2019fdetection}, and Ganomaly~\cite{akcay2019ganomaly} learn only from images of healthy subjects. In addition, we analyzed Brainomaly's performance in transductive and inductive learning settings, provided an ablation study of the object functions, and compared our proposed AUCp metric with FID (\sectionautorefname~\ref{subsec:results_abl}).

All of our models operate on 2D MRI slices. We used 2D sagittal slices for all experiments. We performed a central crop to remove empty regions outside the brain, resulting in $192\times192$ sagittal slices for both datasets. We used $\lambda_{id}$ = 1 and a batch size of 16. We trained the models for 400,000 iterations and saved a model for AUCp calculation after every 10,000 iterations. We have used Adam optimizer with a $1e^{-4}$ learning rate. The learning rate has been decayed for the last 100,000 iterations.

\begin{figure}[!h]
    \centering
    \includegraphics[width=\linewidth]{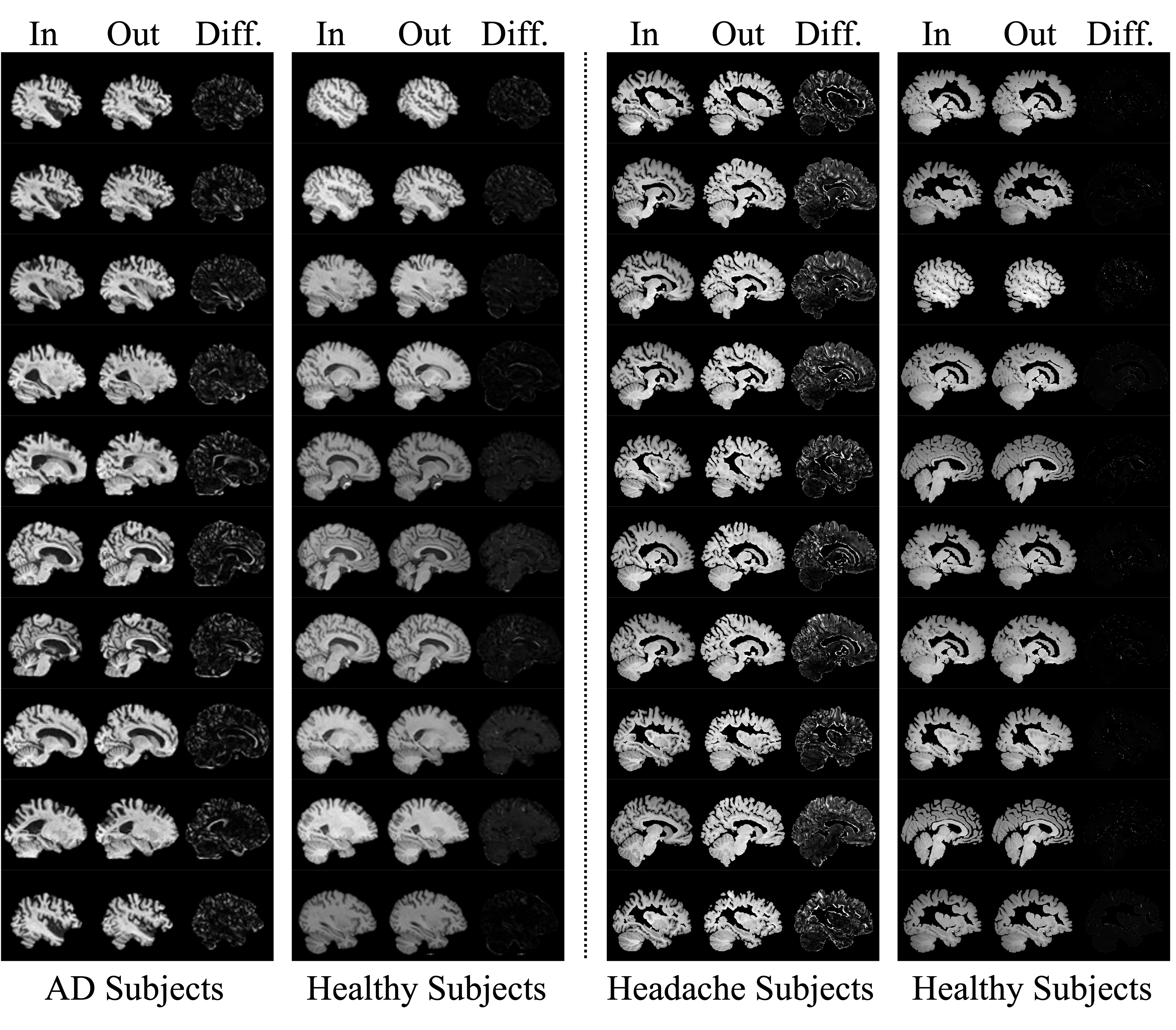}
    \caption{Qualitative results of Alzheimer's disease and headache detection by Brainomaly. The left two columns display the results of Alzheimer's disease detection experiments, while the right two columns depict the outcomes of headache detection experiments. As expected, Brainomaly exhibits higher activations in the difference map for diseased subjects in comparison to healthy subjects, which is the base for its disease detection.}
    \label{fig:qual_result}
\end{figure}

\section{Results and Analyses}
\label{sec:results}

\subsection{Alzheimer's Disease Detection}
\label{subsec:ad_results}

\tableautorefname~\ref{tab:detection_results} compares Brainomaly's Alzheimer's disease detection performance on both {\em AD DS1} and {\em AD DS2} data with six state-of-the-art methods. {\em AD DS1} and {\em AD DS2} columns report the numbers when these data were used as an unseen test set. Brainomlay outperforms the existing methods by a large margin, achieving an average Alzheimer's disease detection AUC of 0.6550. Among the competing methods, f-AnoGAN performed the best. It achieved an average AUC of 0.6020, which is {\em 8.09\% less} than the proposed Brainomaly. Ganomaly and DDAD performed close to f-AnoGAN, achieving average AUC of 0.5956 and 0.5926, respectively. The rest of the competing methods performed like random guesses. \figureautorefname~\ref{fig:qual_result} (left) shows Brainomaly's qualitative results for Alzheimer's disease detection. As seen, the difference maps for subjects with Alzheimer's disease have higher activation than that of healthy subjects. Receiver operating characteristics (ROC) curve analyses are provided in~\appendixautorefname~\ref{appx:roc}.

\subsection{Headache Detection}
\label{subsec:headache_results}

\tableautorefname~\ref{tab:detection_results} also compares Brainomaly's headache detection performance on both {\em HEAD DS1} and {\em HEAD DS2} data with six state-of-the-art methods. Like Alzheimer's disease detection, Brainomaly also outperforms the competing methods in headache detection by a large margin. It achieved an average headache detection AUC of 0.8960. Performing {\em 13.84\% less} than Brainomaly, HealthyGAN achieved the second-best average AUC of 0.7720. Other baseline methods like ALAD, Ganomaly, and DDAD achieved even poorer AUCs of 0.0.7653, 0.6913, and 0.6280, respectively. f-AnoGAN and ALOOC just failed in this task.

Brainomaly also performed better in detecting headache sub-types. On {\em HEAD DS1}, it achieved a precision of 0.9375 (3 incorrect out of 48) in migraine detection, 0.3750 (15 incorrect out of 24; see discussion) in APTH detection, and 0.9600 (only 1 incorrect out of 25) in PPTH detection. On {\em HEAD DS2}, it achieved a precision of 0.9167 (4 incorrect out of 48) in migraine detection, 0.6667 (8 incorrect out of 24) in APTH detection, and 0.9583 (only 1 incorrect out of 24) in PPTH detection.

\figureautorefname~\ref{fig:qual_result} (right) shows Brainomaly's qualitative results for headache detection. Similar to Alzheimer's disease detection, the difference maps for subjects with headaches have higher activation than those for healthy subjects. ROC curve analyses are provided in~\appendixautorefname~\ref{appx:roc}.

\begin{table*}[t!]
    \begin{subtable}[h]{0.48\textwidth}
        \centering
        \caption{Transductive vs. Inductive Learning}
        \begin{tabular}{p{2.5em} c c c c}
            \hline
            \multicolumn{5}{c}{\textbf{Alzheimer's Disease Dataset}} \\
            \hline
             & {\em AD DS1} & {\em AD DS2} & Avg. & {\em p}-value \\
            \hline
            Transduc. & 0.6526 & 0.6825 & 0.6676 & \multirow{2}{*}{0.555} \\
            Inductive & 0.6452 & 0.6648 & 0.6550 \\
            \hline
            \hline
            \multicolumn{5}{c}{\textbf{Headache Dataset}} \\
            \hline
             & {\em HEAD DS1} & {\em HEAD DS2} & Avg. & {\em p}-value \\
            \hline
            Transduc. & 0.9182 & 0.8633 & 0.8908 & \multirow{2}{*}{0.873} \\
            Inductive & 0.9041 & 0.8878 & 0.8960 \\
            \hline
        \end{tabular}
        \label{tab:results_abl_trans_induct}
    \end{subtable}
    \hfill
    \setcounter{subtable}{2}
    \begin{subtable}[!h]{0.48\textwidth}
        \caption{FID vs. AUCp: Correlation with AUC}
        \centering
        \begin{tabular}{l c c}
            \hline
            & \multicolumn{2}{c}{\textbf{Alzheimer's Disease Dataset}} \\
            \hline
             & {\em AD DS1} & {\em AD DS2} \\
            \hline
            FID & 0.5701 & 0.4773 \\
            \textbf{AUCp (Our)} & \textbf{0.9583} & \textbf{0.9656} \\
            \hline
            \hline
            & \multicolumn{2}{c}{\textbf{Headache Dataset}} \\
            \hline
             & {\em HEAD DS1} & {\em HEAD DS2} \\
            \hline
            FID & 0.5227 & 0.3187 \\
            \textbf{AUCp (Our)} & \textbf{0.9528} & \textbf{0.5986} \\
            \hline
        \end{tabular}
        \label{tab:results_abl_fid_aucp_corr}
    \end{subtable}
    \hfill
    \setcounter{subtable}{1}
    \begin{subtable}[h]{0.48\textwidth}
        \centering
        \caption{Importance of Identity Loss}
        \begin{tabular}{l l c c}
            \hline
            \multicolumn{4}{c}{\textbf{Alzheimer's Disease Dataset}} \\
            \hline
             & & {\em AD DS1} & {\em AD DS2} \\
            \hline
            \multirow{2}{*}{Transduc.} & \textbf{Brainomaly} & \textbf{0.6526} & \textbf{0.6825} \\
             & $- \mathcal{L}_{id}$ & 0.6303 & 0.6815 \\
            \hline
            \multirow{2}{*}{Inductive} & \textbf{Brainomaly} & 0.6452 & \textbf{0.6648} \\
             & $- \mathcal{L}_{id}$ & \textbf{0.6521} & 0.6455 \\
            \hline
            \hline
            \multicolumn{4}{c}{\textbf{Headache Dataset}} \\
            \hline
             & & {\em HEAD DS1} & {\em HEAD DS2} \\
            \hline
            \multirow{2}{*}{Transduc.} & \textbf{Brainomaly} & \textbf{0.9182} & \textbf{0.8633} \\
             & $- \mathcal{L}_{id}$ & 0.7824 & 0.8091 \\
            \hline
            \multirow{2}{*}{Inductive} & \textbf{Brainomaly} & \textbf{0.9041} & \textbf{0.8878} \\
             & $- \mathcal{L}_{id}$ & 0.8073 & 0.8359 \\
            \hline
        \end{tabular}
        \label{tab:results_abl_indentity_loss}
    \end{subtable}
    \hfill
    \setcounter{subtable}{3}
    \begin{subtable}[!h]{0.48\textwidth}
        \caption{FID vs. AUCp: Detection Performance}
        \centering
        \begin{tabular}{l l c c}
            \hline
            \multicolumn{4}{c}{\textbf{Alzheimer's Disease Dataset}} \\
            \hline
             & & {\em AD DS1} & {\em AD DS2} \\
            \hline
            \multirow{2}{*}{Transduc.} & FID & 0.618 & 0.6771 \\
             & \textbf{AUCp (Our)} & \textbf{0.6526} & \textbf{0.6825} \\
            \hline
            \multirow{2}{*}{Inductive} & FID & 0.6389 & 0.6453 \\
             & \textbf{AUCp (Our)} & \textbf{0.6452} & \textbf{0.6648} \\
            \hline
            \hline
            \multicolumn{4}{c}{\textbf{Headache Dataset}} \\
            \hline
             & & {\em HEAD DS1} & {\em HEAD DS2} \\
            \hline
            \multirow{2}{*}{Transduc.} & FID & 0.8807 & \textbf{0.9120} \\
             & \textbf{AUCp (Our)} & \textbf{0.9182} & 0.8633 \\
            \hline
            \multirow{2}{*}{Inductive} & FID & 0.9002 & 0.8589 \\
             & \textbf{AUCp (Our)} & \textbf{0.9041} & \textbf{0.8878} \\
            \hline
        \end{tabular}
        \label{tab:results_abl_fid_aucp_detection}
    \end{subtable}
    \caption{Summary of the ablation studies of different components of Brainomaly. These ablation studies show Brainomaly's (a) generalization ability on both unannotated seen and unseen datasets, (b) effectiveness of the objective function, and (c--d) superiority of the proposed AUCp metric for inference model selection.}
\end{table*}

\subsection{Ablation Studies}
\label{subsec:results_abl}

\noindent\textbf{Comparison of Image-to-Image Translation.} To better understand the contribution of Brainomlay's additive map-based image-to-image translation, we have compared it with HealthyGAN~\cite{rahman2022healthygan} by keeping the network architecture, data-split, and inference model selection metrics the same. The results summarized in~\tableautorefname~\ref{tab:detection_results} show that Brainomaly consistently outperformed HealthyGAN across tasks irrespective of inference model selection metrics.

\noindent\textbf{Transductive vs. Inductive Learning.} Using an unannotated set of mixed brain MRIs (\figureautorefname~\ref{fig:anomaly_detection_overview}) allows Brainomaly to operate in transductive and inductive learning modes. Therefore, we evaluate our proposed Brainomaly in both learning settings. For the transductive learning setting, we evaluate Alzheimer's disease and headache detection on the unannotated mixed brain MRI set used during training. In contrast, for the inductive learning setting, we utilize an additional unseen test set for Alzheimer's disease detection and headache detection evaluation. Please note that the performance reported in~\tableautorefname~\ref{tab:detection_results} and analyzed in the previous two subsections were in inductive settings.

\tableautorefname~\ref{tab:results_abl_trans_induct} summarizes Brainomaly's performance for Alzheimer's disease and headache detection in both transductive and inductive learning settings. The average performance of Brainomaly for Alzheimer's disease and headache detection is statistically the same ($p$-value $>$ 0.005) in both transductive and inductive learning settings. These results show that Brainomaly generalizes well on unseen test data.

\noindent\textbf{Impact of Identity Loss on Objective Function.} Inspired from~\cite{siddiquee2019learning}, we incorporated the {\em identity loss} ($\mathcal{L}_{id}$ in~\equationautorefname~\ref{eq:id_loss}) to balance the image translation. As seen in~\tableautorefname~\ref{tab:results_abl_indentity_loss}, $\mathcal{L}_{id}$ plays an important role in Brainomaly's image translation and significantly improves its performance for both Alzheimer's disease and headache detection.

\noindent\textbf{Inference Model Selection---FID~vs.~AUCp.} In~\tableautorefname~\ref{tab:results_abl_fid_aucp_corr}, we have shown that our AUCp score proposed in~\sectionautorefname~\ref{subsec:method_validation} has a stronger correlation with the actual (when all the annotations are available) AUC scores. Therefore, the proposed AUCp metric renders itself a better metric than FID for selecting the model for inference. Please note that~\tableautorefname~\ref{tab:results_abl_fid_aucp_corr} reports absolute correlation values. To further validate, we have provided the AUC scores obtained by the best models according to FID and the AUCp scores in both transductive and inductive learning settings for each dataset in~\tableautorefname~\ref{tab:results_abl_fid_aucp_detection}. It is evident from the figure that the models selected by our AUCp metric dominate in detection performance over the models selected by FID.

\section{Discussion}

The proposed Brainomaly method aims to perform patient-level neurologic disease detection without requiring brain image annotation. Though it generates the difference maps showing structural changes in Alzheimer's disease and headache subjects (\figureautorefname~\ref{fig:qual_result}), these maps are not precise. They show more structural changes than actual changes performed by the underlying diseases; as a result, they are not useful for precise localization. If needed, weakly-supervised localization methods such as GradCAM \cite{selvaraju2017grad}, Fixed-Point GAN~\cite{siddiquee2019learning}, and VAGAN~\cite{baumgartner2017visual} can be utilized for better localization using the patient-level detections from Brainomaly as weak annotations.

The proposed AUCp metric does not guarantee the selection of the best possible model for inference as it uses {\em imperfect} annotations. However, our empirical analyses show that AUCp generally selects a better inference model than the popular FID metric.

In~\sectionautorefname~\ref{subsec:headache_results}, we have seen that APTH detection using Brainomaly is not as good as detecting other headache sub-types. This might be due to the acuity of the condition and greater heterogeneity in brain structural changes amongst these individuals compared to those who have had longstanding migraine or PTH (i.e., those with PPTH). Among the 15 misclassified APTH subjects in {\em HEAD DS1}, we found 5 were recovered at a 3-month time point. This improves the APTH detection rate from 0.3750 to 0.5833. Similarly, in {\em HEAD DS2}, 1 out of 8 misclassified subjects recovered at a 3-month time point, improving the detection rate from 0.6667 to 0.7083. Future studies are needed to explore the heterogeneity amongst those with APTH.

\section{Conclusion}

In conclusion, the proposed unsupervised neurologic disease detection method, Brainomaly, is highly effective in detecting Alzheimer's disease and headaches from T1-weighted brain MRIs, outperforming existing state-of-the-art methods by a large margin. This performance is attributed to Brainomaly's additive map-based image translation, the capability of utilizing unannotated mixed brain MRIs, and better inference model selection using the proposed AUCp metric. Using an unannotated set of mixed brain MRIs enables Brainomaly to operate in both transductive and inductive learning modes, providing flexibility in its application. In addition, we have shown in~\tableautorefname~\ref{tab:detection_results} that the AUCp can select better models even for existing methods, for example, HealthyGAN. We believe the proposed Brainomaly method can be generalized for unsupervised disease detection from other organs and modalities, which we aim to study in our future work.

\noindent\textbf{Acknowledgments.} This research has been supported by the United States Department of Defense W81XWH-15-1-0286 and W81XWH1910534, National Institutes of Health K23NS070891, National Institutes of Health - National Institute of Neurological Disorders and Stroke, Award Number 1R61NS113315–01, and Amgen Investigator Sponsored Study 20187183. We thank Arizona State University Research Computing (ASURC) for hosting and maintaining our computing resources.

\newpage

{\small
\bibliographystyle{ieee_fullname}
\bibliography{egbib}

\begin{thebibliography}{10}\itemsep=-1pt

\bibitem{akcay2019ganomaly}
Samet Akcay, Amir Atapour-Abarghouei, and Toby~P Breckon.
\newblock Ganomaly: Semi-supervised anomaly detection via adversarial training.
\newblock In {\em Computer Vision--ACCV 2018: 14th Asian Conference on Computer
  Vision, Perth, Australia, December 2--6, 2018, Revised Selected Papers, Part
  III 14}, pages 622--637. Springer, 2019.

\bibitem{alami2018unsupervised}
Youssef Alami~Mejjati, Christian Richardt, James Tompkin, Darren Cosker, and
  Kwang~In Kim.
\newblock Unsupervised attention-guided image-to-image translation.
\newblock {\em Advances in neural information processing systems}, 31, 2018.

\bibitem{alex2017generative}
Varghese Alex, Mohammed~Safwan KP, Sai~Saketh Chennamsetty, and Ganapathy
  Krishnamurthi.
\newblock Generative adversarial networks for brain lesion detection.
\newblock In {\em Medical Imaging 2017: Image Processing}, volume 10133, pages
  113--121. SPIE, 2017.

\bibitem{arjovsky2017wasserstein}
Martin Arjovsky, Soumith Chintala, and L{\'e}on Bottou.
\newblock Wasserstein generative adversarial networks.
\newblock In {\em International conference on machine learning}, pages
  214--223. PMLR, 2017.

\bibitem{bai2022novel}
Tian Bai, Mingyu Du, Lin Zhang, Lei Ren, Li Ruan, Yuan Yang, Guanghao Qian,
  Zihao Meng, Li Zhao, and M~Jamal Deen.
\newblock A novel alzheimer’s disease detection approach using gan-based
  brain slice image enhancement.
\newblock {\em Neurocomputing}, 492:353--369, 2022.

\bibitem{baumgartner2017visual}
Christian~F Baumgartner, Lisa~M Koch, Kerem~Can Tezcan, Jia~Xi Ang, and Ender
  Konukoglu.
\newblock Visual feature attribution using wasserstein gans.
\newblock In {\em Proc IEEE Comput Soc Conf Comput Vis Pattern Recognit}, 2017.

\bibitem{baydargil2021anomaly}
Husnu~Baris Baydargil, Jang-Sik Park, and Do-Young Kang.
\newblock Anomaly analysis of alzheimer’s disease in pet images using an
  unsupervised adversarial deep learning model.
\newblock {\em Applied Sciences}, 11(5):2187, 2021.

\bibitem{ixi}
Ixi dataset, NA.

\bibitem{cabreza2022anomaly}
Jean~Nathan Cabreza, Geoffrey~A Solano, Sun~Arthur Ojeda, and Vincent Munar.
\newblock Anomaly detection for alzheimer’s disease in brain mris via
  unsupervised generative adversarial learning.
\newblock In {\em 2022 International Conference on Artificial Intelligence in
  Information and Communication (ICAIIC)}, pages 1--5. IEEE, 2022.

\bibitem{cai2022dual}
Yu Cai, Hao Chen, Xin Yang, Yu Zhou, and Kwang-Ting Cheng.
\newblock Dual-distribution discrepancy for anomaly detection in chest x-rays.
\newblock {\em arXiv preprint arXiv:2206.03935}, 2022.

\bibitem{chen2018unsuperviseddetection}
Xiaoran Chen and Ender Konukoglu.
\newblock Unsupervised detection of lesions in brain {MRI} using constrained
  adversarial auto-encoders.
\newblock {\em CoRR}, abs/1806.04972, 2018.

\bibitem{choi2019deep}
Hongyoon Choi, Seunggyun Ha, Hyejin Kang, Hyekyoung Lee, Dong~Soo Lee,
  Alzheimer's Disease~Neuroimaging Initiative, et~al.
\newblock Deep learning only by normal brain pet identify unheralded brain
  anomalies.
\newblock {\em EBioMedicine}, 43:447--453, 2019.

\bibitem{choi2018stargan}
Yunjey Choi, Minje Choi, Munyoung Kim, Jung-Woo Ha, Sunghun Kim, and Jaegul
  Choo.
\newblock Stargan: Unified generative adversarial networks for multi-domain
  image-to-image translation.
\newblock In {\em Proceedings of the IEEE conference on computer vision and
  pattern recognition}, pages 8789--8797, 2018.

\bibitem{choi2020stargan}
Yunjey Choi, Youngjung Uh, Jaejun Yoo, and Jung-Woo Ha.
\newblock Stargan v2: Diverse image synthesis for multiple domains.
\newblock In {\em Proceedings of the IEEE/CVF Conference on Computer Vision and
  Pattern Recognition}, pages 8188--8197, 2020.

\bibitem{donahue2016adversarialdetection}
Jeff Donahue, Philipp Kr{\"{a}}henb{\"{u}}hl, and Trevor Darrell.
\newblock Adversarial feature learning.
\newblock {\em CoRR}, abs/1605.09782, 2016.

\bibitem{esteva2017dermatologist}
Andre Esteva, Brett Kuprel, Roberto~A Novoa, Justin Ko, Susan~M Swetter,
  Helen~M Blau, and Sebastian Thrun.
\newblock Dermatologist-level classification of skin cancer with deep neural
  networks.
\newblock {\em nature}, 542(7639):115--118, 2017.

\bibitem{gherbi2019encodingdetection}
Elies Gherbi, Blaise Hanczar, Jean-Christophe Janodet, and Witold Klaudel.
\newblock An encoding adversarial network for anomaly detection.
\newblock In {\em Asian Conference on Machine Learning}, pages 188--203. PMLR,
  2019.

\bibitem{goodfellow2020generative}
Ian Goodfellow, Jean Pouget-Abadie, Mehdi Mirza, Bing Xu, David Warde-Farley,
  Sherjil Ozair, Aaron Courville, and Yoshua Bengio.
\newblock Generative adversarial networks.
\newblock {\em Communications of the ACM}, 63(11):139--144, 2020.

\bibitem{goodfellow2014generative}
Ian~J Goodfellow, Jean Pouget-Abadie, Mehdi Mirza, Bing Xu, David Warde-Farley,
  Sherjil Ozair, Aaron~C Courville, and Yoshua Bengio.
\newblock Generative adversarial nets.
\newblock In {\em NIPS}, 2014.

\bibitem{gulrajani2017improved}
Ishaan Gulrajani, Faruk Ahmed, Martin Arjovsky, Vincent Dumoulin, and Aaron~C
  Courville.
\newblock Improved training of wasserstein gans.
\newblock {\em Advances in neural information processing systems}, 30, 2017.

\bibitem{han2021madgan}
Changhee Han, Leonardo Rundo, Kohei Murao, Tomoyuki Noguchi, Yuki Shimahara,
  Zolt{\'a}n~{\'A}d{\'a}m Milacski, Saori Koshino, Evis Sala, Hideki Nakayama,
  and Shin’ichi Satoh.
\newblock Madgan: Unsupervised medical anomaly detection gan using multiple
  adjacent brain mri slice reconstruction.
\newblock {\em BMC bioinformatics}, 22(2):1--20, 2021.

\bibitem{he2015delving}
Kaiming He, Xiangyu Zhang, Shaoqing Ren, and Jian Sun.
\newblock Delving deep into rectifiers: Surpassing human-level performance on
  imagenet classification.
\newblock In {\em Proceedings of the IEEE international conference on computer
  vision}, pages 1026--1034, 2015.

\bibitem{he2019attgantrans}
Zhenliang He, Wangmeng Zuo, Meina Kan, Shiguang Shan, and Xilin Chen.
\newblock Arbitrary facial attribute editing: Only change what you want.
\newblock {\em CoRR}, abs/1711.10678, 2017.

\bibitem{heusel2017gans}
Martin Heusel, Hubert Ramsauer, Thomas Unterthiner, Bernhard Nessler, and Sepp
  Hochreiter.
\newblock Gans trained by a two time-scale update rule converge to a local nash
  equilibrium.
\newblock {\em Advances in neural information processing systems}, 30, 2017.

\bibitem{isola2017image}
Phillip Isola, Jun-Yan Zhu, Tinghui Zhou, and Alexei~A Efros.
\newblock Image-to-image translation with conditional adversarial networks.
\newblock In {\em Proceedings of the IEEE conference on computer vision and
  pattern recognition}, pages 1125--1134, 2017.

\bibitem{jin2021unsupervised}
Shichen Jin, Peini Zou, Ying Han, and Jiehui Jiang.
\newblock Unsupervised detection of individual atrophy in alzheimer's disease.
\newblock In {\em 2021 43rd Annual International Conference of the IEEE
  Engineering in Medicine \& Biology Society (EMBC)}, pages 2647--2650. IEEE,
  2021.

\bibitem{kim2017learning}
Taeksoo Kim, Moonsu Cha, Hyunsoo Kim, Jung~Kwon Lee, and Jiwon Kim.
\newblock Learning to discover cross-domain relations with generative
  adversarial networks.
\newblock In {\em International conference on machine learning}, pages
  1857--1865. PMLR, 2017.

\bibitem{ledig2017photo}
Christian Ledig, Lucas Theis, Ferenc Husz{\'a}r, Jose Caballero, Andrew
  Cunningham, Alejandro Acosta, Andrew Aitken, Alykhan Tejani, Johannes Totz,
  Zehan Wang, et~al.
\newblock Photo-realistic single image super-resolution using a generative
  adversarial network.
\newblock In {\em Proceedings of the IEEE conference on computer vision and
  pattern recognition}, pages 4681--4690, 2017.

\bibitem{li2016precomputed}
Chuan Li and Michael Wand.
\newblock Precomputed real-time texture synthesis with markovian generative
  adversarial networks.
\newblock In {\em European conference on computer vision}, pages 702--716.
  Springer, 2016.

\bibitem{liu2019stgan}
Ming Liu, Yukang Ding, Min Xia, Xiao Liu, Errui Ding, Wangmeng Zuo, and Shilei
  Wen.
\newblock Stgan: A unified selective transfer network for arbitrary image
  attribute editing.
\newblock In {\em Proceedings of the IEEE/CVF conference on computer vision and
  pattern recognition}, pages 3673--3682, 2019.

\bibitem{liu2017unsupervised}
Ming-Yu Liu, Thomas Breuel, and Jan Kautz.
\newblock Unsupervised image-to-image translation networks.
\newblock {\em Advances in neural information processing systems}, 30, 2017.

\bibitem{nizan2020breaking}
Ori Nizan and Ayellet Tal.
\newblock Breaking the cycle-colleagues are all you need.
\newblock In {\em Proceedings of the IEEE/CVF conference on computer vision and
  pattern recognition}, pages 7860--7869, 2020.

\bibitem{rahman2023headache}
Md~Mahfuzur Rahman~Siddiquee, Jay Shah, Catherine Chong, Simona Nikolova, Gina
  Dumkrieger, Baoxin Li, Teresa Wu, and Todd~J Schwedt.
\newblock Headache classification and automatic biomarker extraction from
  structural mris using deep learning.
\newblock {\em Brain Communications}, 5(1):fcac311, 2023.

\bibitem{rahman2022healthygan}
Md~Mahfuzur Rahman~Siddiquee, Jay Shah, Teresa Wu, Catherine Chong, Todd
  Schwedt, and Baoxin Li.
\newblock Healthygan: Learning from unannotated medical images to detect
  anomalies associated with human disease.
\newblock In {\em International Workshop on Simulation and Synthesis in Medical
  Imaging}, pages 43--54. Springer, 2022.

\bibitem{siddiquee2019learning}
Md~Mahfuzur Rahman~Siddiquee, Zongwei Zhou, Nima Tajbakhsh, Ruibin Feng,
  Michael~B Gotway, Yoshua Bengio, and Jianming Liang.
\newblock Learning fixed points in generative adversarial networks: From
  image-to-image translation to disease detection and localization.
\newblock In {\em Proceedings of the IEEE/CVF international conference on
  computer vision}, pages 191--200, 2019.

\bibitem{sabokrou2018adversariallydetection}
Mohammad Sabokrou, Mohammad Khalooei, Mahmood Fathy, and Ehsan Adeli.
\newblock Adversarially learned one-class classifier for novelty detection.
\newblock {\em CoRR}, abs/1802.09088, 2018.

\bibitem{schlegl2017unsupervised}
Thomas Schlegl, Philipp Seeb{\"o}ck, Sebastian~M Waldstein, Ursula
  Schmidt-Erfurth, and Georg Langs.
\newblock Unsupervised anomaly detection with generative adversarial networks
  to guide marker discovery.
\newblock In {\em Information Processing in Medical Imaging: 25th International
  Conference, IPMI 2017, Boone, NC, USA, June 25-30, 2017, Proceedings}, pages
  146--157. Springer, 2017.

\bibitem{schlegl2019fdetection}
Thomas Schlegl, Philipp Seeböck, Sebastian~M. Waldstein, Georg Langs, and
  Ursula Schmidt-Erfurth.
\newblock f-anogan: Fast unsupervised anomaly detection with generative
  adversarial networks.
\newblock {\em Medical Image Analysis}, 54:30--44, 2019.

\bibitem{selvaraju2017grad}
Ramprasaath~R Selvaraju, Michael Cogswell, Abhishek Das, Ramakrishna Vedantam,
  Devi Parikh, and Dhruv Batra.
\newblock Grad-cam: Visual explanations from deep networks via gradient-based
  localization.
\newblock In {\em Proceedings of the IEEE International Conference on Computer
  Vision}, pages 618--626, 2017.

\bibitem{shen2017learning}
Wei Shen and Rujie Liu.
\newblock Learning residual images for face attribute manipulation.
\newblock In {\em Proceedings of the IEEE conference on computer vision and
  pattern recognition}, pages 4030--4038, 2017.

\bibitem{yang2018multimodal}
Hao Yang, Junran Zhang, Qihong Liu, and Yi Wang.
\newblock Multimodal mri-based classification of migraine: using deep learning
  convolutional neural network.
\newblock {\em Biomedical engineering online}, 17(1):1--14, 2018.

\bibitem{yi2017dualgan}
Zili Yi, Hao Zhang, Ping Tan, and Minglun Gong.
\newblock Dualgan: Unsupervised dual learning for image-to-image translation.
\newblock In {\em Proceedings of the IEEE international conference on computer
  vision}, pages 2849--2857, 2017.

\bibitem{zenati2018efficientdetection}
Houssam Zenati, Chuan~Sheng Foo, Bruno Lecouat, Gaurav Manek, and
  Vijay~Ramaseshan Chandrasekhar.
\newblock Efficient gan-based anomaly detection.
\newblock {\em arXiv preprint arXiv:1802.06222}, 2018.

\bibitem{zenati2018adversarially}
Houssam Zenati, Manon Romain, Chuan-Sheng Foo, Bruno Lecouat, and Vijay
  Chandrasekhar.
\newblock Adversarially learned anomaly detection.
\newblock In {\em 2018 IEEE International conference on data mining (ICDM)},
  pages 727--736. IEEE, 2018.

\bibitem{zhang2018generativetrans}
Gang Zhang, Meina Kan, Shiguang Shan, and Xilin Chen.
\newblock Generative adversarial network with spatial attention for face
  attribute editing.
\newblock In {\em Proceedings of the European Conference on Computer Vision
  (ECCV)}, September 2018.

\bibitem{zhao2020unpaired}
Yihao Zhao, Ruihai Wu, and Hao Dong.
\newblock Unpaired image-to-image translation using adversarial consistency
  loss.
\newblock In {\em Computer Vision--ECCV 2020: 16th European Conference,
  Glasgow, UK, August 23--28, 2020, Proceedings, Part IX 16}, pages 800--815.
  Springer, 2020.

\bibitem{zhu2017unpaired}
Jun-Yan Zhu, Taesung Park, Phillip Isola, and Alexei~A Efros.
\newblock Unpaired image-to-image translation using cycle-consistent
  adversarial networks.
\newblock In {\em Proceedings of the IEEE international conference on computer
  vision}, pages 2223--2232, 2017.

\bibitem{zhu2017toward}
Jun-Yan Zhu, Richard Zhang, Deepak Pathak, Trevor Darrell, Alexei~A Efros,
  Oliver Wang, and Eli Shechtman.
\newblock Toward multimodal image-to-image translation.
\newblock {\em Advances in neural information processing systems}, 30, 2017.

\end{thebibliography}
}

\newpage
\onecolumn
\appendix

\section{Supplementary Materials}

\subsection{AUCp Metric Calculation}
\label{appx:aucp}

\begin{figure}[h!]
    \centering
    \includegraphics[width=\textwidth]{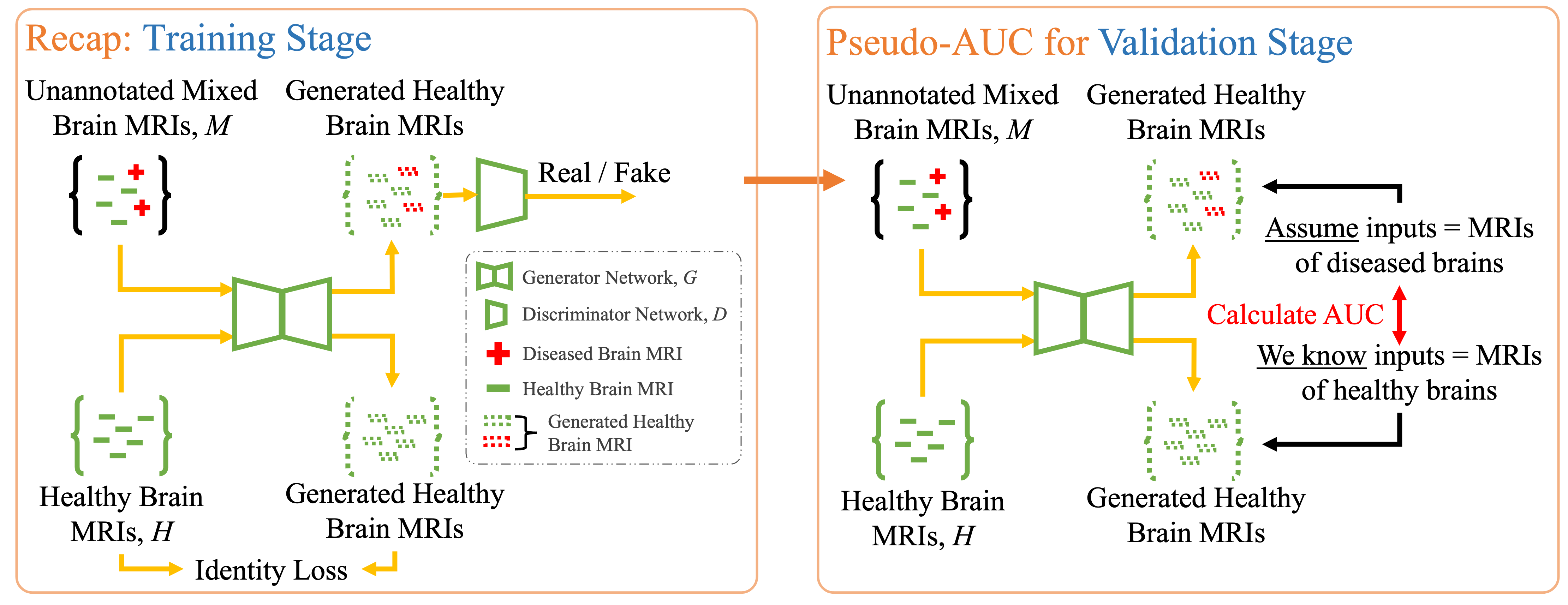}
    \caption{An overview of the proposed AUCp metric calculation. To calculate AUCp, we first generate the {\em disease detection scores} for each model, as discussed in~\sectionautorefname~\ref{subsec:method_disease_detection}. As we already know that the set $H$ contains only healthy-brain MRIs, we assume the labels for MRIs in the unannotated mixed brain MRI set, $M$, to be diseased brains. Then, we use these {\em imperfect} annotations as ground truths along with the {\em disease detection scores} in the traditional AUC calculation resulting in AUCp scores. Once the AUCp scores are available for all the models, we select a model with the highest AUCp score for inference. In~\sectionautorefname~\ref{subsec:results_abl}, we have also shown that AUCp sets a better-performing model for inference compared to FID, commonly used in existing works~\cite{rahman2022healthygan}.}
    \label{fig:aucp_calculation}
\end{figure}

\subsection{Neural Network Architectures}
\label{appx:arch}

\noindent A detailed neural network architecture for the discriminator is provided in~\tableautorefname~\ref{tab:arch_discriminator} and for the generator is provided in~\tableautorefname~\ref{tab:arch_generator}. Both these networks are adopted from~\cite{choi2018stargan,siddiquee2019learning,rahman2022healthygan} with slight modification.

\begin{table}[!htp]\centering
    \bgroup
    \def\arraystretch{1.5}
    \begin{tabular}{|c|p{6cm}|c|c|}
        \hline
        \textbf{Type} & \textbf{Operations} & \textbf{Input Shape} & \textbf{Output Shape} \\
        \hline
        Input layer & Conv2d (OC=64, KS=4, S=2, P=1), LeakyReLU (NS=0.01) & $(h, w, 3)$ & $(\frac{h}{2}, \frac{w}{2}, 64)$ \\
        \hline
        \multirow{5}{*}{Hidden layers} & Conv2d (OC=128, KS=4, S=2, P=1), LeakyReLU (NS=0.01) & $(\frac{h}{2}, \frac{w}{2}, 64)$ & $(\frac{h}{4}, \frac{w}{4}, 128)$ \\
        & Conv2d (OC=256, KS=4, S=2, P=1), LeakyReLU (NS=0.01)  & $(\frac{h}{4}, \frac{w}{4}, 128)$ & $(\frac{h}{8}, \frac{w}{8}, 256)$ \\
        & Conv2d (OC=512, KS=4, S=2, P=1), LeakyReLU (NS=0.01)  & $(\frac{h}{8}, \frac{w}{8}, 256)$ & $(\frac{h}{16}, \frac{w}{16}, 512)$ \\
        & Conv2d (OC=1024, KS=4, S=2, P=1), LeakyReLU (NS=0.01) & $(\frac{h}{16}, \frac{w}{16}, 512)$ & $(\frac{h}{32}, \frac{w}{32}, 1024)$ \\
        & Conv2d (OC=2048, KS=4, S=2, P=1), LeakyReLU (NS=0.01) & $(\frac{h}{32}, \frac{w}{32}, 1024)$ & $(\frac{h}{64}, \frac{w}{64}, 2048)$ \\
        \hline
        Output layer ($D_{src}$) & Conv2d (OC=1, KS=3, S=1, P=1) & $(\frac{h}{64}, \frac{w}{64}, 2048)$ & $(\frac{h}{64}, \frac{w}{64}, 1)$ \\ \hline
    \end{tabular}
    \egroup
\caption{Discriminator network architecture. OC, KS, S, P, and NS stand for output channels, kernel size, stride, padding, and negative slope, respectively.}
\label{tab:arch_discriminator}
\end{table}

\begin{table}[!htp]\centering
    \bgroup
    \def\arraystretch{1.5}
    \begin{tabular}{|c|p{8cm}|c|c|}
        \hline
        \textbf{Type} & \textbf{Operations} & \textbf{Input Shape} & \textbf{Output Shape} \\
        \hline
        \multirow{3}{*}{Encoder} & Conv2d (OC=64, KS=7, S=1, P=3), IN, ReLU & $(h, w, 3)$ & $(h, w, 64)$ \\
        & Conv2d (OC=128, KS=4, S=2, P=1), IN, ReLU & $(h, w, 64)$ & $(\frac{h}{2}, \frac{w}{2}, 128)$ \\
        & Conv2d (OC=256, KS=4, S=2, P=1), IN, ReLU & $(\frac{h}{2}, \frac{w}{2}, 128)$ & $(\frac{h}{4}, \frac{w}{4}, 256)$ \\
        \hline
        \multirow{6}{*}{Bottleneck} & Residual Block: Conv2d (OC=256, KS=3, S=1, P=1), IN, ReLU & $(\frac{h}{4}, \frac{w}{4}, 256)$ & $(\frac{h}{4}, \frac{w}{4}, 256)$ \\
        & Residual Block: Conv2d (OC=256, KS=3, S=1, P=1), IN, ReLU & $(\frac{h}{4}, \frac{w}{4}, 256)$ & $(\frac{h}{4}, \frac{w}{4}, 256)$ \\
        & Residual Block: Conv2d (OC=256, KS=3, S=1, P=1), IN, ReLU & $(\frac{h}{4}, \frac{w}{4}, 256)$ & $(\frac{h}{4}, \frac{w}{4}, 256)$ \\
        & Residual Block: Conv2d (OC=256, KS=3, S=1, P=1), IN, ReLU & $(\frac{h}{4}, \frac{w}{4}, 256)$ & $(\frac{h}{4}, \frac{w}{4}, 256)$ \\
        & Residual Block: Conv2d (OC=256, KS=3, S=1, P=1), IN, ReLU & $(\frac{h}{4}, \frac{w}{4}, 256)$ & $(\frac{h}{4}, \frac{w}{4}, 256)$ \\
        & Residual Block: Conv2d (OC=256, KS=3, S=1, P=1), IN, ReLU & $(\frac{h}{4}, \frac{w}{4}, 256)$ & $(\frac{h}{4}, \frac{w}{4}, 256)$ \\
        \hline
        \multirow{3}{*}{Decoder} & ConvTranspose2d (OC=128, KS=4, S=2, P=1), IN, ReLU & $(\frac{h}{4}, \frac{w}{4}, 256)$ & $(\frac{h}{4}, \frac{w}{4}, 128)$ \\
        & ConvTranspose2d (OC=64, KS=4, S=2, P=1), IN, ReLU & $(\frac{h}{2}, \frac{w}{2}, 128)$ & $(h, w, 64)$ \\
        & ConvTranspose2d (OC=1, KS=7, S=1, P=3) & $(h, w, 64)$ & $(h, w, 1)$ \\
        \hline
    \end{tabular}
    \egroup
\caption{Generator network architecture. OC, KS, S, P, and IN stand for output channels, kernel size, stride, padding, and instance norm, respectively.}
\label{tab:arch_generator}
\end{table}

\clearpage

\subsection{Receiver Operating Characteristic Curves}
\label{appx:roc}

\begin{figure}[!htp]
\begin{tikzpicture}
\begin{axis}[
  xmin=0,
  xmax=1,
  ymin=0,
  ymax=1,
  xtick={0, 0.1, 0.2, 0.3, 0.4, 0.5, 0.6, 0.7, 0.8, 0.9, 1},
  ytick={0, 0.1, 0.2, 0.3, 0.4, 0.5, 0.6, 0.7, 0.8, 0.9, 1},
  xmajorgrids=true,
  ymajorgrids=true,
  grid style=dashed,
  legend columns=2,
  legend style={at={(0.44,-0.2)},anchor=north},
  legend cell align=left,
  domain=0:1,
  ignore zero=y,
  title={AD DS1},
  xlabel={False Positive Rate},
  ylabel={True Positive Rate},
]

\addplot [line width=0.5mm, red] table {roc_data/ad_ds1_brainomaly.dat};
\addplot [line width=0.5mm, brown!50] table {roc_data/ad_ds1_fanogan.dat};
\addplot [line width=0.5mm, cyan!50] table {roc_data/ad_ds1_ddad.dat};
\addplot [line width=0.5mm, violet!50] table {roc_data/ad_ds1_ganomaly.dat};
\addplot [line width=0.5mm, green!50] table {roc_data/ad_ds1_alad.dat};
\addplot [line width=0.5mm, orange!50] table {roc_data/ad_ds1_alocc.dat};
\addplot [line width=0.5mm, blue!50] table {roc_data/ad_ds1_healthygan.dat};
\addplot [line width=0.5mm, black, dashed] table {roc_data/line.dat};

\legend{Brainomaly (0.6452), f-AnoGAN (0.5946), DDAD (0.5897), Ganomaly (0.5864), ALAD (0.5329), ALOCC (0.4670), HealthyGAN (0.4598)}

\end{axis}
\end{tikzpicture}
\begin{tikzpicture}
\begin{axis}[
  xmin=0,
  xmax=1,
  ymin=0,
  ymax=1,
  xtick={0, 0.1, 0.2, 0.3, 0.4, 0.5, 0.6, 0.7, 0.8, 0.9, 1},
  ytick={0, 0.1, 0.2, 0.3, 0.4, 0.5, 0.6, 0.7, 0.8, 0.9, 1},
  xmajorgrids=true,
  ymajorgrids=true,
  grid style=dashed,
  legend columns=2,
  legend style={at={(0.44,-0.2)},anchor=north},
  legend cell align=left,
  domain=0:1,
  ignore zero=y,
  title={HEAD DS1},
  xlabel={False Positive Rate},
  ylabel={True Positive Rate},
]

\addplot [line width=0.5mm, red] table {roc_data/head_ds1_brainomaly.dat};
\addplot [line width=0.5mm, green!50] table {roc_data/head_ds1_alad.dat};
\addplot [line width=0.5mm, violet!50] table {roc_data/head_ds1_ganomaly.dat};
\addplot [line width=0.5mm, blue!50] table {roc_data/head_ds1_healthygan.dat};
\addplot [line width=0.5mm, cyan!50] table {roc_data/head_ds1_ddad.dat};
\addplot [line width=0.5mm, brown!50] table {roc_data/head_ds1_fanogan.dat};
\addplot [line width=0.5mm, orange!50] table {roc_data/head_ds1_alocc.dat};
\addplot [line width=0.5mm, black, dashed] table {roc_data/line.dat};

\legend{Brainomaly (0.9041), ALAD (0.7819), Ganomaly (0.7313), HealthyGAN (0.7107), DDAD (0.6128), f-AnoGAN (0.4354), ALOCC (0.3044)}

\end{axis}
\end{tikzpicture}

\begin{tikzpicture}
\begin{axis}[
  xmin=0,
  xmax=1,
  ymin=0,
  ymax=1,
  xtick={0, 0.1, 0.2, 0.3, 0.4, 0.5, 0.6, 0.7, 0.8, 0.9, 1},
  ytick={0, 0.1, 0.2, 0.3, 0.4, 0.5, 0.6, 0.7, 0.8, 0.9, 1},
  xmajorgrids=true,
  ymajorgrids=true,
  grid style=dashed,
  legend columns=2,
  legend style={at={(0.44,-0.2)},anchor=north},
  legend cell align=left,
  domain=0:1,
  ignore zero=y,
  title={AD DS2},
  xlabel={False Positive Rate},
  ylabel={True Positive Rate},
]

\addplot [line width=0.5mm, red] table {roc_data/ad_ds2_brainomaly.dat};
\addplot [line width=0.5mm, brown!50] table {roc_data/ad_ds2_fanogan.dat};
\addplot [line width=0.5mm, violet!50] table {roc_data/ad_ds2_ganomaly.dat};
\addplot [line width=0.5mm, cyan!50] table {roc_data/ad_ds2_ddad.dat};
\addplot [line width=0.5mm, blue!50] table {roc_data/ad_ds2_healthygan.dat};
\addplot [line width=0.5mm, green!50] table {roc_data/ad_ds2_alad.dat};
\addplot [line width=0.5mm, orange!50] table {roc_data/ad_ds2_alocc.dat};
\addplot [line width=0.5mm, black, dashed] table {roc_data/line.dat};

\legend{Brainomaly (0.6648), f-AnoGAN (0.6093), Ganomaly (0.6048), DDAD (0.5955), HealthyGAN (0.5468), ALAD (0.5239), ALOCC (0.4746)}

\end{axis}
\end{tikzpicture}
\begin{tikzpicture}
\begin{axis}[
  xmin=0,
  xmax=1,
  ymin=0,
  ymax=1,
  xtick={0, 0.1, 0.2, 0.3, 0.4, 0.5, 0.6, 0.7, 0.8, 0.9, 1},
  ytick={0, 0.1, 0.2, 0.3, 0.4, 0.5, 0.6, 0.7, 0.8, 0.9, 1},
  xmajorgrids=true,
  ymajorgrids=true,
  grid style=dashed,
  legend columns=2,
  legend style={at={(0.44,-0.2)},anchor=north},
  legend cell align=left,
  domain=0:1,
  ignore zero=y,
  title={HEAD DS2},
  xlabel={False Positive Rate},
  ylabel={True Positive Rate},
]

\addplot [line width=0.5mm, red] table {roc_data/head_ds2_brainomaly.dat};
\addplot [line width=0.5mm, blue!50] table {roc_data/head_ds2_healthygan.dat};
\addplot [line width=0.5mm, green!50] table {roc_data/head_ds2_alad.dat};
\addplot [line width=0.5mm, orange!50] table {roc_data/head_ds2_alocc.dat};
\addplot [line width=0.5mm, violet!50] table {roc_data/head_ds2_ganomaly.dat};
\addplot [line width=0.5mm, cyan!50] table {roc_data/head_ds2_ddad.dat};
\addplot [line width=0.5mm, brown!50] table {roc_data/head_ds2_fanogan.dat};
\addplot [line width=0.5mm, black, dashed] table {roc_data/line.dat};

\legend{Brainomaly (0.8878), HealthyGAN (0.8333), ALAD (0.7486), ALOCC (0.6566), Ganomaly (0.6514), DDAD (0.6431), f-AnoGAN (0.3925)}

\end{axis}
\end{tikzpicture}

\caption{Receiver operative characteristics analyses for Alzheimer's disease and headache detection. {\em AD DS1} and {\em AD DS2} on the left compares Alzheimer's disease detection performances and {\em HEAD DS1} and {\em HEAD DS2} on the right compares headache detection performances. As discussed in~\sectionautorefname~\ref{sec:results}, Brainomaly outperforms the existing methods across the tasks.}
\label{appx_fig:roc}
\end{figure}
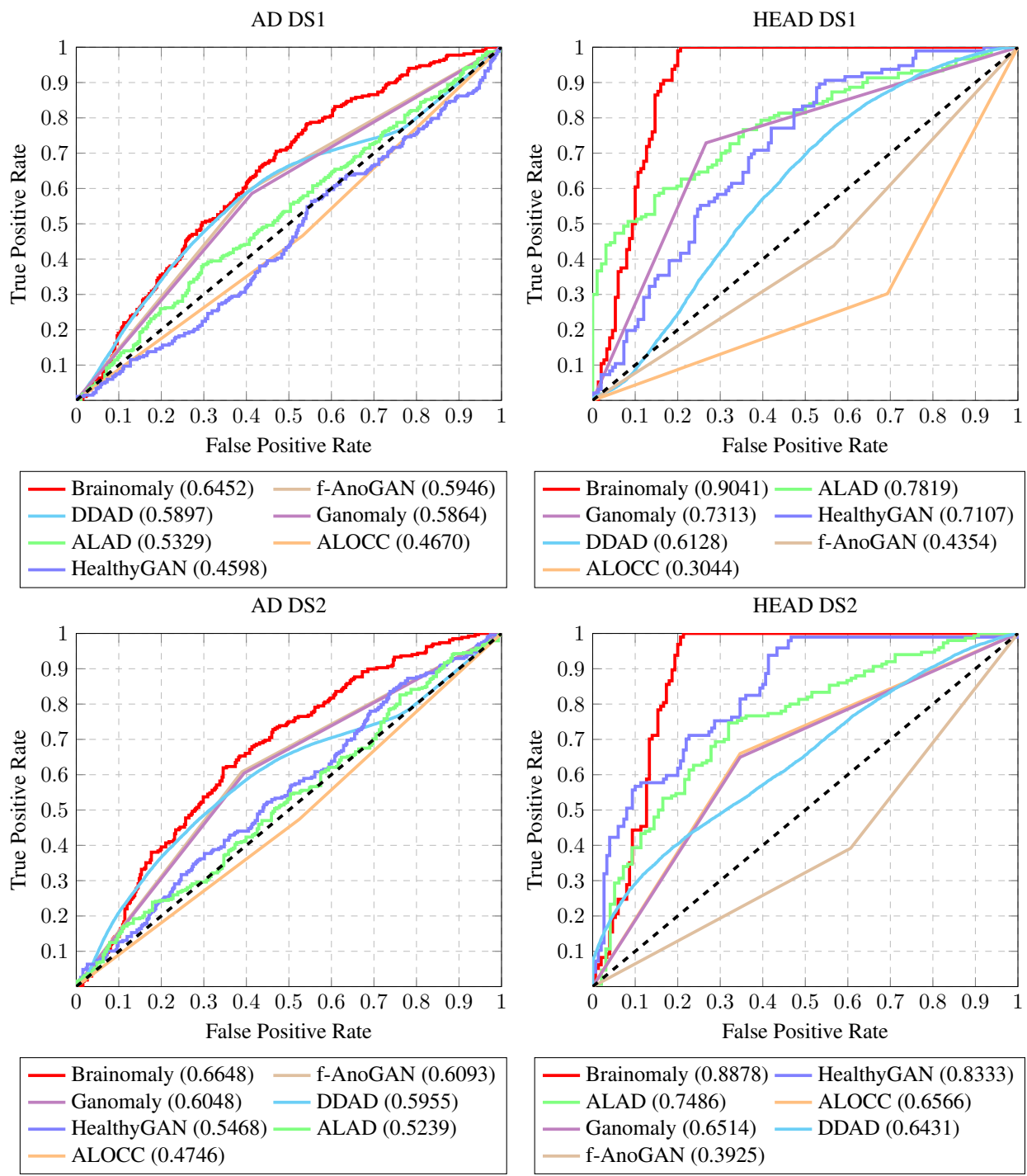

\end{document}